\documentclass[showpacs,preprintnumbers,amsmath,amssymb]{revtex4}
\usepackage{amsmath,amssymb,graphics,epsfig,subfigure}
\usepackage{color}

\begin{document}
\renewcommand{\baselinestretch}{1.3}

\title{Geodesics and periodic orbits in Kehagias-Sfetsos black holes in deformed Ho\u{r}ava-Lifshitz gravity}

\author{Shao-Wen Wei$^{1,2}$ \footnote{weishw@lzu.edu.cn},
        Jie Yang$^{1}$ \footnote{yangjiev@lzu.edu.cn},
        Yu-Xiao Liu$^{1}$ \footnote{liuyx@lzu.edu.cn}}

\affiliation{$^{1}$Institute of Theoretical Physics $\&$ Research Center of Gravitation, Lanzhou University, Lanzhou 730000, People's Republic of China\\
$^{2}$Department of Physics and Astronomy, University of Waterloo, Waterloo, Ontario, Canada, N2L 3G1}

\begin{abstract}
The motion of a massive test particle around a Kehagias-Sfetsos black hole in deformed Ho\u{r}ava-Lifshitz gravity is studied. Employing the effective potential, the marginally bound orbits and the innermost stable circular orbits are analyzed. For the marginally bound orbits, their radius and angular momentum decrease with the parameter $\omega$ of the gravity. For the innermost stable circular orbits, the energy and angular momentum also decrease with $\omega$. Based on these results, we investigate the periodic orbits in the Kehagias-Sfetsos black holes. It is found that the apsidal angle parameter increases with the particle energy, while decreases with the angular momentum. Moreover, compared to the Schwarzschild black hole, the periodic orbits in Kehagias-Sfetsos black holes always have lower energy. These results provide us a possible way to distinguish the Kehagias-Sfetsos black hole in deformed Ho\u{r}ava-Lifshitz gravity from the Schwarzschild black hole.
\end{abstract}



\maketitle

\section{Introduction}

Black holes are the most fascinating objects in our Universe. Due to the strong gravity, there are many interesting astronomical phenomena around them, such as the black hole lensing, shadow, and accelerator. In particular, with the help of the geodesics of a test particle, these phenomena can be well studied. Therefore, investigating different orbital trajectories for photons and massive particles is of great interest.

Recently, motivated by the detection of gravitational waves \cite{Abbott}, the study of the geodesics has received extensive attention. For example, in the ringdown stage of the black hole merger, the frequency of the gravitational waves is characterized by the linearized vibrational modes, the quasinormal modes. In the eikonal limit, such mode is associated with the spacetime's light ring \cite{Mashhoona,Schutz,IyerWill,Cardoso}, which can be obtained by the analysis of the null geodesics. On the other hand, the final spin of the black hole can also be well estimated by the Buonanno-Kidder-Lehner receipt \cite{Buonanno} through solving the innermost stable circular orbits (ISCOs) from the time-like geodesics. During the inspiral stage, accompanied by the gravitational wave radiation, the two initial black holes of extreme mass ratio approach to each other. In this process, periodic orbits act as successive orbit transition states and play an important role in the study of the gravitational wave radiation \cite{Glampedakis} . Inspired by this, Levin \emph{et. al.} proposed a classification for the periodic orbits of massive particles, which is very useful for understanding the dynamics of the black hole merger. In their classification, each periodic orbit is characterized by three integers, $z$, $w$, and $v$, which, respectively, denote the zoom, whirl, and vertex behaviors. This study has been carried out for the Schwarzschild black holes, Kerr black holes, charged black holes, Kerr-Sen black holes, and naked singularities \cite{Levin,Grossman,Levin2,Misra,Babar,LiuDing}.

On the other hand, supermassive black hole surrounded by a test particle is reminiscent of a hydrogen atom while in a macroscopic level. Then we can make an analogy between these periodic orbits and the ones of the electron around an atom. Thus, we can obtain the discrete energy levels for the black hole system \cite{Levin2}. By measuring these energy levels, we can extract the information of the black hole. Especially, combining with the black hole thermodynamics, we can explore the area and entropy spectra for the black hole \cite{Bekenstein1}. Therefore, with this energy levels, the fundamental black hole physics will be revealed.

From another side, modified gravity continues to be an important and fascinating subject in gravitational physics. Among them, Ho\u{r}ava-Lifshitz (HL) theory \cite{Horava} proposed by Ho\u{r}ava at a Lifshitz point is an interesting one. It keeps spatial general covariance and time reparametrization invariance. Such theory also performs as a good candidate for studying the quantum field theory of gravity. In the infrared divergence limit, it can reduce to the Einstein's gravity. A progress report on the recent developments of HL gravity can be found in Ref. \cite{anzhong}. In this paper, we aim to study the geodesics and periodic orbits in the black hole in deformed HL gravity, and to show the difference between it and general relativity. This will also bring us the insight in detecting the gravitational waves from the black hole merger in deformed HL gravity.

The present paper is organized as follows. In Sec. \ref{bhg}, we first introduce the black hole solution and its geodesics. Then in Sec. \ref{epbo}, by employing the effective potential, we investigate the marginally bound orbits and ISCOs. Then, based on the result, the periodic orbits are studied in detail. Finally, we summarize and discuss our results in Sec. \ref{Conclusion}.

\section{Black hole and geodesics}
\label{bhg}

In the limit of cosmological constant $\Lambda_W \to 0$, the action of the deformed HL gravity is given by \cite{KS}
\begin{eqnarray}
S_{HL}&=&\int dtd^3x \Big({\cal L}_0 + {\cal L}_1\Big),
\end{eqnarray}
where ${\cal L}_0$ and ${\cal L}_1$ read
\begin{eqnarray}
 {\cal L}_0 &=& \sqrt{g}N\left\{\frac{2}{\kappa^2}(K_{ij}K^{ij}
  -\lambda K^2)+\frac{\kappa^2\mu^2(\Lambda_W R
  -3\Lambda_W^2)}{8(1-3\lambda)}\right\}\,,\\
 {\cal L}_1&=&
\sqrt{g}N\left\{\frac{\kappa^2\mu^2 (1-4\lambda)}{32(1-3\lambda)}R^2
-\frac{\kappa^2}{2\tilde{\omega}^4} \left(C_{ij} -\frac{\mu \tilde{\omega}^2}{2}R_{ij}\right)
\left(C^{ij} -\frac{\mu \tilde{\omega}^2}{2}R^{ij}\right) +\mu^4R
\right\}.
\end{eqnarray}
Here $\tilde{\omega}$, $\lambda$, $\mu$, and $\kappa$ are parameters in HL gravity, and the extrinsic curvature $K_{ij}$ and Cotton tensor $C_{ij}$ are
\begin{eqnarray}
 K_{ij} &=& \frac{1}{2N} \Bigg(\partial_t g_{ij}-\nabla_i N_j -
\nabla_j N_i\Bigg),\\
 C^{ij}&=&\epsilon^{ik\ell}\nabla_k\left(R^j{}_\ell
-\frac14R\delta_\ell^j\right)\nonumber\\
   &=&\epsilon^{ik\ell}\nabla_k R^j{}_\ell
-\frac14\epsilon^{ikj}\partial_kR.
\end{eqnarray}
For $\lambda=1$ or $\tilde{\omega}=16\mu^{2}/\kappa^{2}$, there is a static and
asymptotically flat black hole solution given by Kehagias and Sfetsos (KS) \cite{KS}
\begin{eqnarray}
 ds^{2}=-N^{2}(r)dt^{2}+\frac{1}{f(r)}dr^{2}+r^{2}(d\theta^{2}+\sin^{2}\theta d\phi^{2}),
\end{eqnarray}
where the metric functions are given by
\begin{eqnarray}
 N^{2}(r)=f(r)=1+\tilde{\omega}r^{2}-\sqrt{\tilde{\omega}^{2}r^{4}+4\tilde{\omega}Mr}.
\end{eqnarray}
When $\tilde{\omega}\rightarrow\infty$, such black hole will reduce to the Schwarzschild black hole. In order to clearly show their difference, we make a parameter transformation,
\begin{eqnarray}
 \omega=\frac{1}{2\tilde{\omega}^{2}}.
\end{eqnarray}
Thus the parameter $\omega$ has a positive value, and the Schwarzschild black hole will be recovered with $\omega$=0. The metric functions will be of the following form
\begin{eqnarray}
 N^{2}(r)=f(r)=1+\frac{1}{2\omega^{2}}(r^{2}-\sqrt{r^{4}+8Mr\omega^{2}}).
\end{eqnarray}
Solving $f(r)=0$, we can obtain the horizon radius for the KS black hole: 
\begin{eqnarray}
 r_{\pm}=M\pm\sqrt{M^{2}-\omega^{2}}.
\end{eqnarray}
A KS black hole corresponds to $\omega/M\leq1$. While the extremal black hole occurs at $\omega/M=1$, where the two horizons coincide with each other.

Now let us turn to the geodesics. For a particle freely moving in a spherically symmetric black hole background, its motion is always in the equatorial plane under an appropriate coordinate system. Thus without loss of generality, we set $\theta=\pi/2$. Then the Lagrangian for a particle is
\begin{eqnarray}
 2\mathcal{L}=g_{\mu\nu}\dot{x}^{\mu}\dot{x}^{\nu}
 =g_{tt}\dot{t}^{2}+g_{rr}\dot{r}^{2}+g_{\phi\phi}\dot{\phi}^{2}
 =\delta,
\end{eqnarray}
with $\delta$=-1 and $0$ for massive and massless particles, respectively. Here we consider a massive particle, so we take $\delta$=-1. Following the Lagrangian, the generalized momentum is $p_{\mu}=\frac{\partial\mathcal{L}}{\partial\dot{x}^{\mu}}=g_{\mu\nu}\dot{x}^{\nu}$. On the other hand, this spacetime admits two Killing fields $\partial_{t}$ and $\partial_{\phi}$. So there are two corresponding constants $E$ and $l$ for each geodesic curve, which are the conservation of energy and orbital angular momentum
per unit mass of the motion. Thus we have
\begin{eqnarray}
 p_{t}&=&g_{tt}\dot{t}=-E,\\
 p_{\phi}&=&g_{\phi\phi}\dot{\phi}=l,\\
 p_{r}&=&g_{rr}\dot{r}.
\end{eqnarray}
Solving the above three equations, we have
\begin{eqnarray}
 \dot{t}&=&\frac{E}{N^{2}(r)},\\
 \dot{\phi}&=&\frac{l^{2}}{r^{2}},\\
 \dot{r}^{2}&=&f(r)\bigg(\frac{E^{2}}{N^{2}(r)}-\frac{l^{2}}{r^{2}}-1\bigg).\label{rr}
\end{eqnarray}
Obviously, such geodesic motion for the KS black hole depends on the parameter $\omega$ of the deformed HL gravity. Note that the study of the geodesics was carried out with different interests in Refs. \cite{ChenJing,ChenWang,Setare,Abdujabbarov,Enolskii,Vieira}.

\section{Effective potential and bound orbits}
\label{epbo}

In this section, we mainly study the $r$-motion. For simplicity, we expression Eq. (\ref{rr}) in the following form
\begin{eqnarray}
 \dot{r}^{2}=E^{2}-V_\text{eff},
\end{eqnarray}
with the effective potential given by
\begin{eqnarray}
 V_{\text{eff}}=f(r)\bigg(1+\frac{l^{2}}{r^{2}}\bigg).\label{effective}
\end{eqnarray}
As $r\rightarrow\infty$, we can expand the metric function $f(r)$
\begin{eqnarray}
 f(r\rightarrow\infty)=1-\frac{2M}{r}+\frac{4M^{2}\omega^{2}}{r^{4}}
    +\mathcal{O}\left(\frac{1}{r^{7}}\right).
\end{eqnarray}
Therefore, one has $V_\text{eff}|_{r\rightarrow\infty}$=1. In general, a bound orbit is bounded by two turning points. One is near the black hole, while other one is far from it. With the increase of the energy $E$, the farther one will get more farther, and it approaches infinity when $E$=1. Above this value, the particle will have positive velocity at infinity, i.e., $\dot{r}^{2}=E^{2}-V_\text{eff}|_{r\rightarrow\infty}>0$, and which will lead to no bound orbit. Therefore, $E=1$ is the maximum of the energy for the bound orbits.

\subsection{Marginally bound orbits}

\begin{figure}
\center{
\includegraphics[width=8cm]{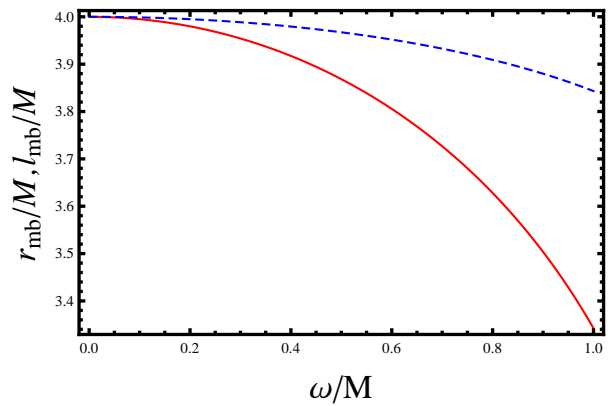}}
\caption{The radius (red solid line) and angular momentum (blue dashed line) for the marginally bound orbits.}\label{Rlw}
\end{figure}

Here, we consider the marginally bound orbits, which are defined by
\begin{eqnarray}
 V_\text{eff}=1,\quad \partial_{r}V_\text{eff}=0.
\end{eqnarray}
Plunging (\ref{effective}) into the above equation, we get
\begin{eqnarray}
 r^{4}+l^{2}r^{2}+2l^{2}\omega^{2}-(l^{2}+r^{2})\sqrt{r^{4}+8Mr\omega^{2}}=0, \label{MarginallyBoundOrbits1}\\
 r^{6}+2Mr^{3}\omega^{2}-6l^{2}Mr\omega^{2}-(r^{4}-2l^{2}\omega^{2})\sqrt{r^{4}+8Mr\omega^{2}}=0.
   \label{MarginallyBoundOrbits2}
\end{eqnarray}
For a fixed $\omega$, we can obtain the radius $r_{\text{mb}}$ and angular momentum $l_{\text{mb}}$ by solving Eqs.~\eqref{MarginallyBoundOrbits1} and \eqref{MarginallyBoundOrbits2}, and the numerical result is given in Fig. \ref{Rlw}. From it, we clearly see that both the radius and angular momentum decrease with $\omega$. And for the extremal black hole ($\omega=M$), we have $r_{\text{mb}}=3.3422M$ and $l_{\text{mb}}=3.8432M$.

\subsection{Innermost stable circular orbits}

\begin{figure}
\center{\subfigure[]{\label{Lisco}
\includegraphics[width=8cm]{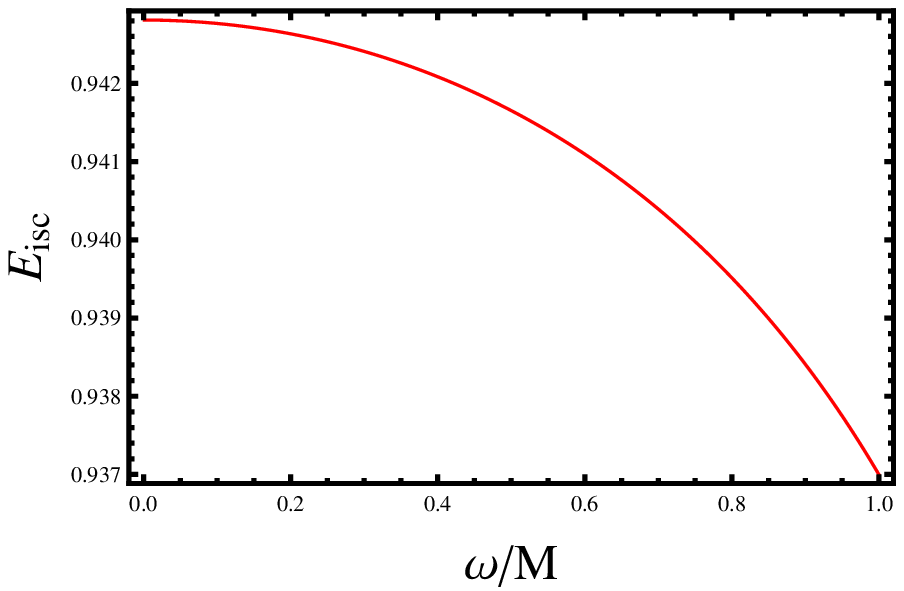}}
\subfigure[]{\label{Eisco}
\includegraphics[width=8cm]{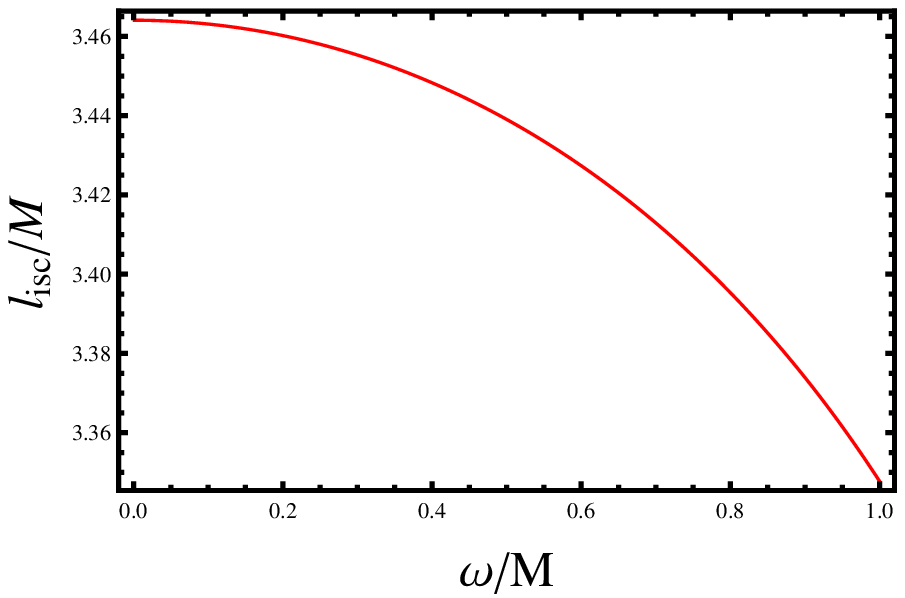}}}
\caption{The energy $E_{\text{isc}}$ and angular momentum $L_{\text{isc}}$ for the ISCOs. (a) $E_{\text{isc}}$ vs $\omega$. (b) $L_{\text{isc}}$ vs $\omega$.}\label{pEisco}
\end{figure}

The ISCOs satisfy the following conditions
\begin{eqnarray}
 V_\text{eff}=E^{2},\quad \partial_{r}V_\text{eff}=0,\quad \partial_{r,r}V_\text{eff}=0,
\end{eqnarray}
from which one can obtain the corresponding energy and momentum:
\begin{eqnarray}
 E_{\text{isc}}&=&\frac{f(r_{\text{isc}})}{\sqrt{f(r_{\text{isc}})-r_{\text{isc}}f'(r_{\text{isc}})/2}},\\
 l_{\text{isc}}&=&r_{\text{isc}}^{3/2}\sqrt{\frac{f'(r_{\text{isc}})}{2f(r_{\text{isc}})-r_{\text{isc}}f'(r_{\text{isc}})}},
\end{eqnarray}
where the radius of the ISCOs can be calculated from the following relation:
\begin{eqnarray}
 r_{\text{isc}}=\frac{3f(r_{\text{isc}})f'(r_{\text{isc}})}{2f'^{2}(r_{\text{isc}})-f(r_{\text{isc}})f''(r_{\text{isc}})}.
\end{eqnarray}
We plot the result in Fig. \ref{pEisco}. It is clear that both the energy and angular momentum for the ISCOs decrease with $\omega$. For the extremal KS black hole, one has $E_{\text{isc}}=0.9370$ and $l_{\text{isc}}=3.3477M$, and the radius of the corresponding orbit is $r_{\text{isc}}=5.2366M$.

\begin{figure}
\center{\subfigure[]{\label{Veff}
\includegraphics[width=8cm]{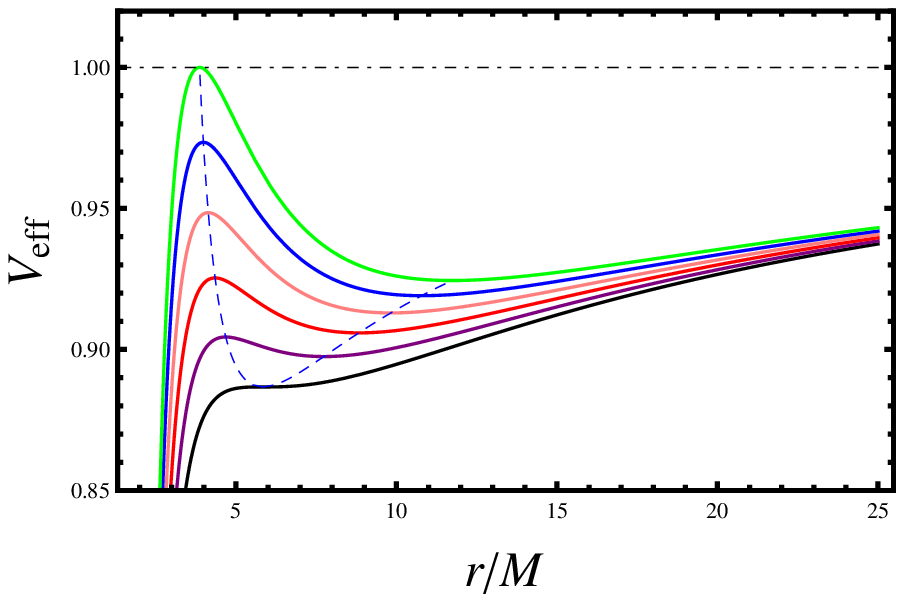}}
\subfigure[]{\label{rr2}
\includegraphics[width=8cm]{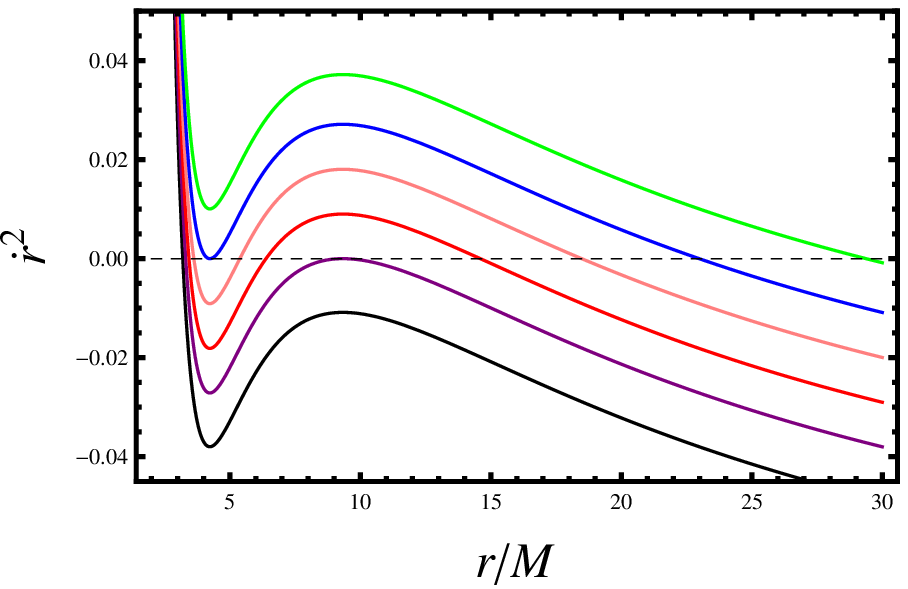}}}
\caption{(a) The effective potential $V_\text{eff}$ as a function of $r/M$ with $\omega$=0.5. The angular momentum $l/M$ varies from $l_{\text{isc}}$ to $l_{\text{mb}}$ from bottom to top. The dashed line represents the extremal points of the effective potential. For this case, the outer black hole horizon locates at $r_{+}=1.87M$. (b) $\dot{r}^{2}$ as a function of $r/M$ with $\omega$=0.5 and $l=\frac{1}{2}(l_{\text{isc}}+l_{\text{mb}})$. The energy $E$ varies from 0.948 to 0.973 from bottom to top.}\label{prr2}
\end{figure}

The behavior of the effective potential is given in Fig. \ref{Veff} with $\omega$=0.5. The angular momentum $l/M$ varies from $l_{\text{isc}}$ to $l_{\text{mb}}$ from bottom to top. The top one describes the case of the marginally bound orbit, which has two extremal points. Decreasing $l/M$, the two points get closer to each other, and finally meet each other at $l=l_{\text{isc}}$ for the ISCOs (bottom one) with the radius $r_{\text{isc}}\approx5.8M$.

\begin{figure}
\center{\subfigure[$\omega$=0.3]{\label{Elma}
\includegraphics[width=8cm]{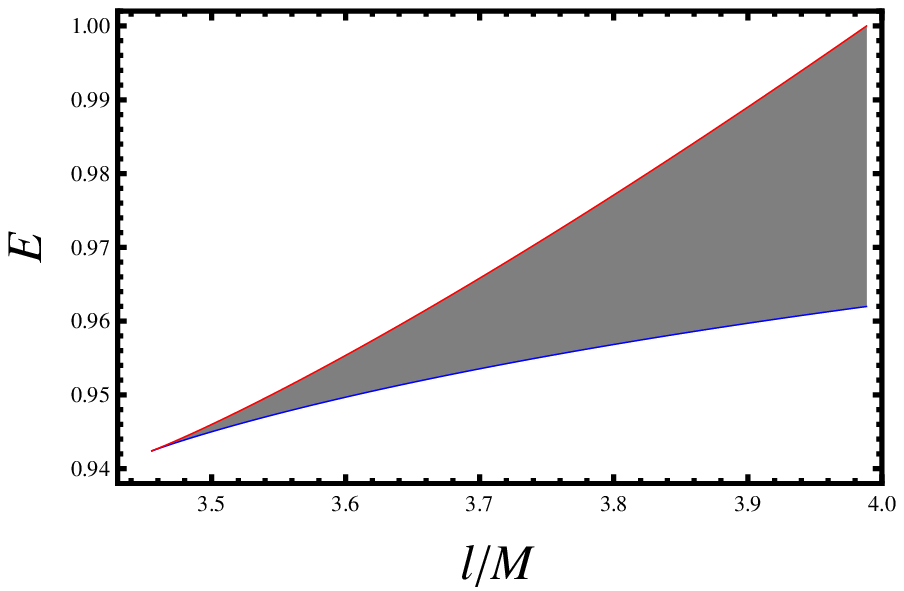}}
\subfigure[$\omega$=0.5]{\label{Elmb}
\includegraphics[width=8cm]{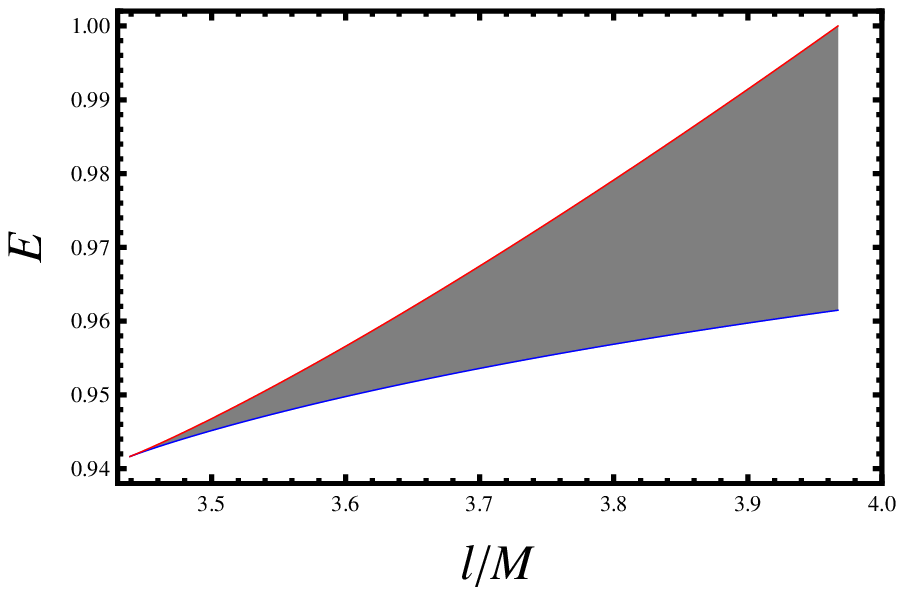}}}
\caption{Parameter regions for the bound orbits (in shadow). (a) $\omega$=0.3. (b) $\omega$=0.5.}\label{pElmb}
\end{figure}

Moreover, for a fixed angular momentum $l=\frac{1}{2}(l_{\text{isc}}+l_{\text{mb}})$, we plot the behavior of $\dot{r}^{2}$ as a function of $r/M$ with $\omega$=0.5 in Fig. \ref{rr2}. From bottom to top, the energy $E$ varies from 0.948 to 0.973. From the figure, we can see that each curve admits two extremal points. If the two points have negative values, these trajectories are forbidden for a massive particle from infinity. On the other hand, if two extremal points are positive, the particle can start at a finite distance, and then fall into the black hole.
A detailed analysis shows that the bound orbits can be only allowed for the case that one extremal point is positive while the other is negative. For example, the energy bound is (0.9537, 0.9678) for $l/M=3.7032$. Further, for each fixed energy $E$, the turning points can be obtained by solving $\dot{r}^{2}=0$. For example, the turning points $r/M$=6.0095, 15.8782 for $E=0.96$. In Fig. \ref{pElmb}, we list the regions for the bound orbits in the $E$-$l/M$ diagram for $\omega$=0.3 and 0.5, respectively. When the parameters fall in the shadow regions, there will be the bound orbits, otherwise, the orbits will not be bounded. Moreover, it is clear that, for a fixed $l/M$, the energy for the bound orbits has a width, and this width increases with $l/M$.

\section{Periodic orbits}

In this section, we would like to examine the periodic orbits around the KS black hole in the HL gravity. According to the viewpoint of Ref. \cite{Levin}, a generic orbit can be treated as a perturbation of periodic orbits. Thus the study of the periodic orbits is very helpful for understanding the nature of any generic orbits and gravitational radiation.

Periodic orbit is one special kind of bound orbit. In this paper, we mainly focus on a spherically symmetric black hole background. So the periodic orbit is completely characterized by the frequencies of oscillations in the $r$-motion  and $\phi$-motion. In particular, a periodic orbit requires that the ratio of these two frequencies must be a rational number such that the particle can return to its initial location after a finite time.

For a bound orbit with two turning points $r_{1}$ and $r_{2}$, the particle is reflected in the range of $(r_{1}, r_{2})$. For each period in $r$, the apsidal angle $\Delta\phi$ passed by the particle is
\begin{eqnarray}
 \Delta\phi=\oint d\phi.
\end{eqnarray}
By making use of the geodesics, the above quantity can be further expressed as
\begin{eqnarray}
 \Delta\phi&=&2{\int_{\phi_{1}}^{\phi_{2}}d\phi} \nonumber\\
           &=&2\int_{r_{1}}^{r_{2}}\frac{\dot{\phi}}{\dot{r}}dr\nonumber\\
           &=&2\int_{r_{1}}^{r_{2}}
           \frac{l}{r^{2}\sqrt{E^{2}-f(r)(1+\frac{l^{2}}{r^{2}})}}dr.\label{knk}
\end{eqnarray}
The factor `2' comes from the symmetry analysis of the geodesics. From this expression, we can clearly see that $\Delta\phi$ closely depends on the energy $E$ and angular momentum $l$, as well as the metric function $f(r)$. Thus, for black holes of different $\omega$, the apsidal angle $\Delta\phi$ will be different. Here we would like to mention that by the inversion of general hyperelliptic integrals of the first, second, and third kind developed in Ref. \cite{Sirimachan}, many properties of the integral of the apsidal angle $\Delta\phi$ can be analytically obtained in some black hole backgrounds, such as one particular black hole in HL gravity and Myers–Perry black hole. However, here it is different to apply this method. So we will numerically solve the integral.

\begin{figure}
\center{
\subfigure[$\epsilon=0.3$]{\label{qe03}
\includegraphics[width=6cm]{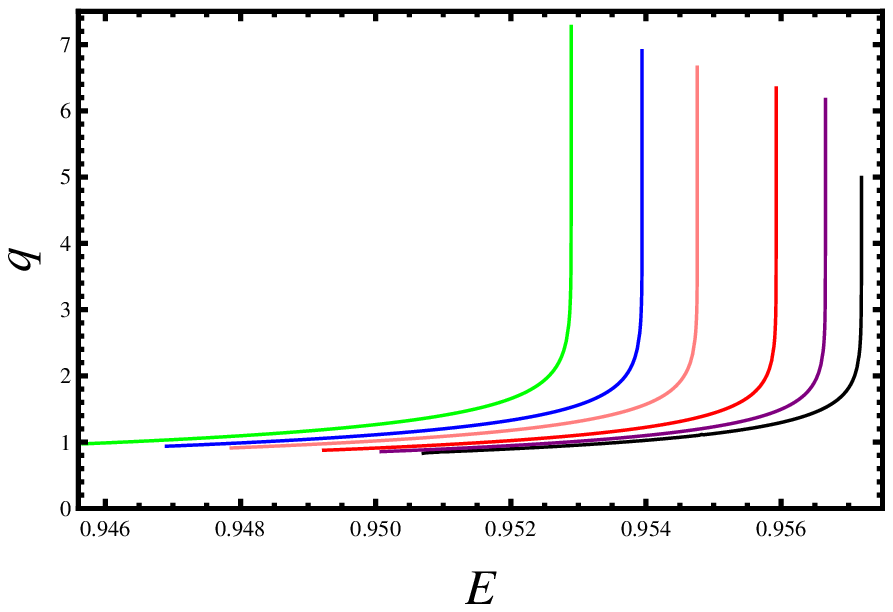}}
\subfigure[$\epsilon=0.5$]{\label{qe05}
\includegraphics[width=6cm]{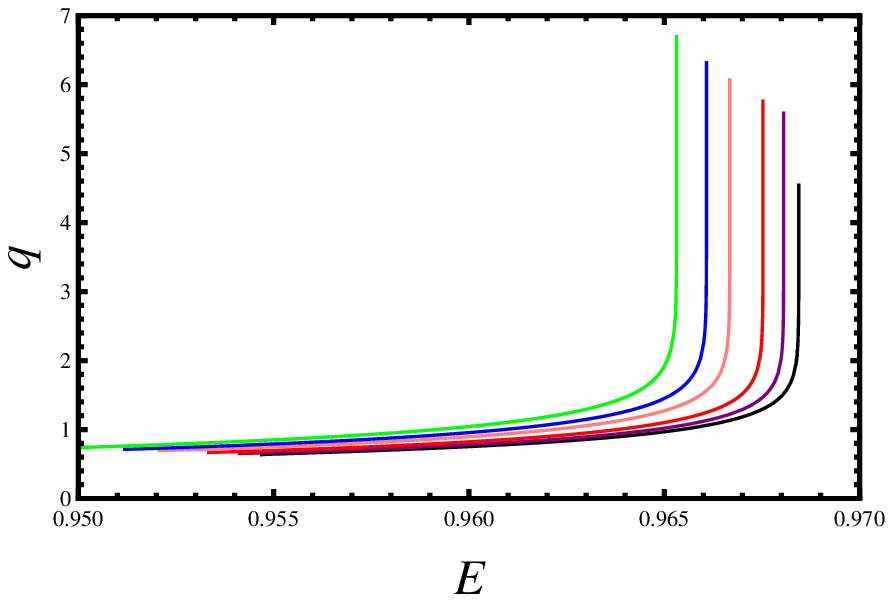}}}\\
\center{\subfigure[$\epsilon=0.7$]{\label{qe07}
\includegraphics[width=6cm]{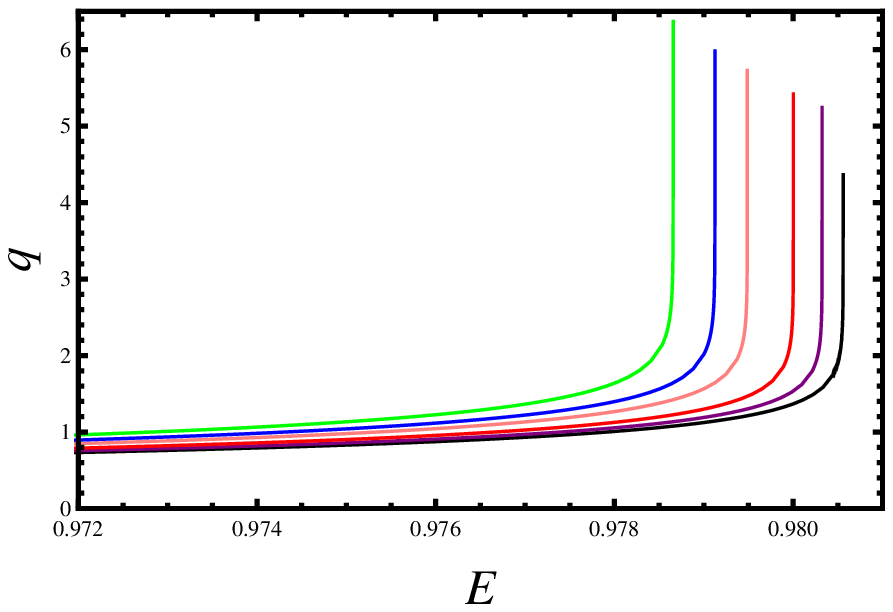}}
\subfigure[$\epsilon=0.9$]{\label{qe09}
\includegraphics[width=6cm]{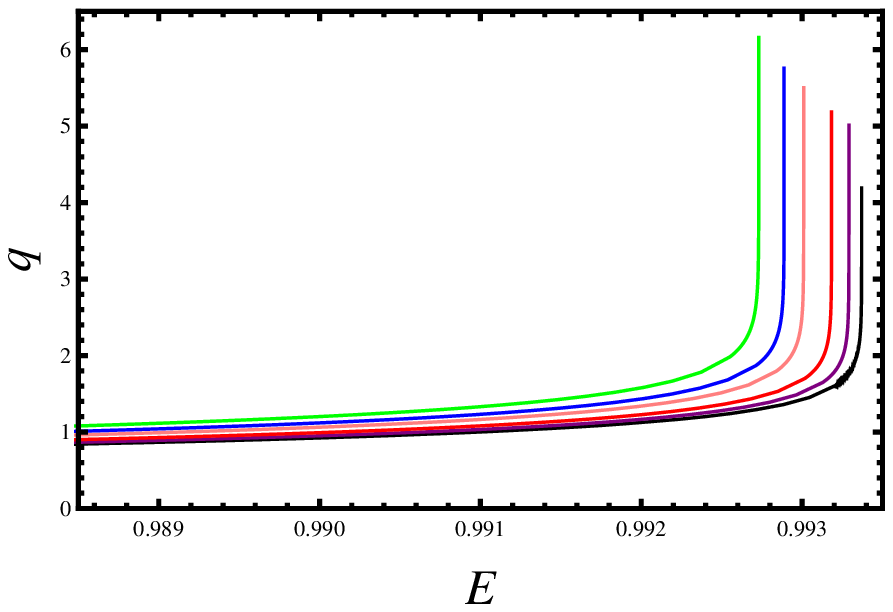}}}
\caption{$q$ vs $E$. The parameter $\omega$=0, 0.4, 0.6, 0.8, 0.9, and 1 from right to left. (a) $\epsilon=0.3$. (b) $\epsilon=0.5$. (c) $\epsilon=0.7$. (d) $\epsilon=0.9$.}\label{pqe09}
\end{figure}

\begin{figure}
\center{\subfigure[$E=0.95$]{\label{ql95}
\includegraphics[width=6cm]{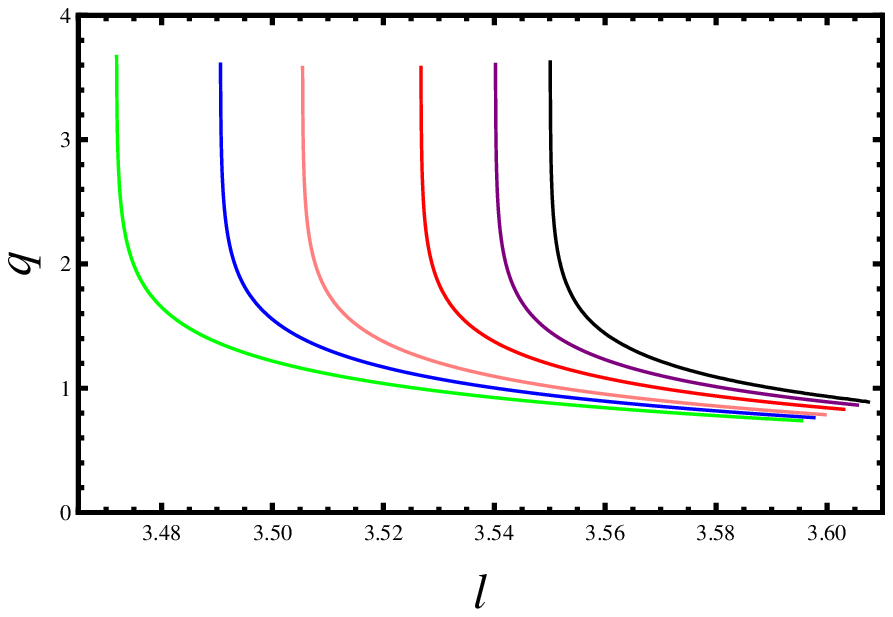}}
\subfigure[$E=0.96$]{\label{ql96}
\includegraphics[width=6cm]{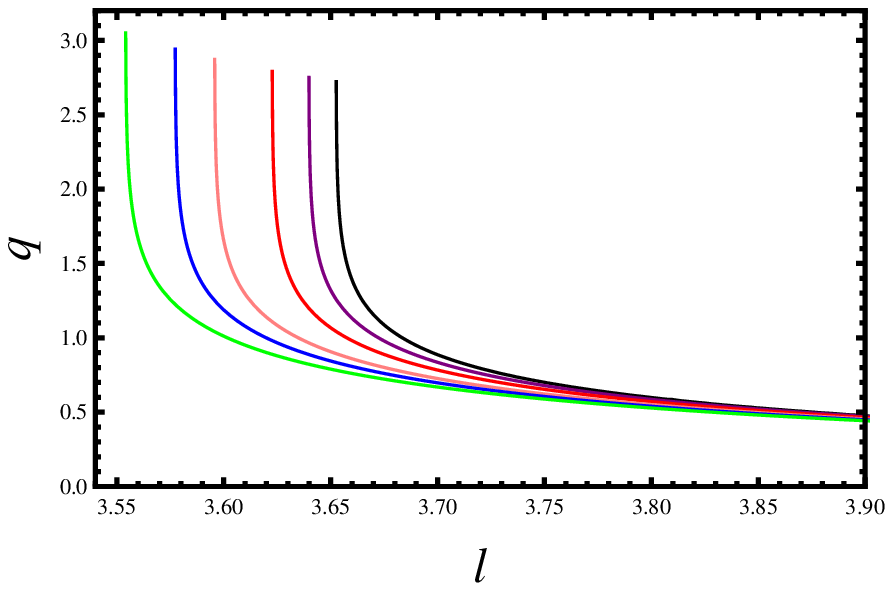}}}\\
\center{\subfigure[$E=0.97$]{\label{ql97}
\includegraphics[width=6cm]{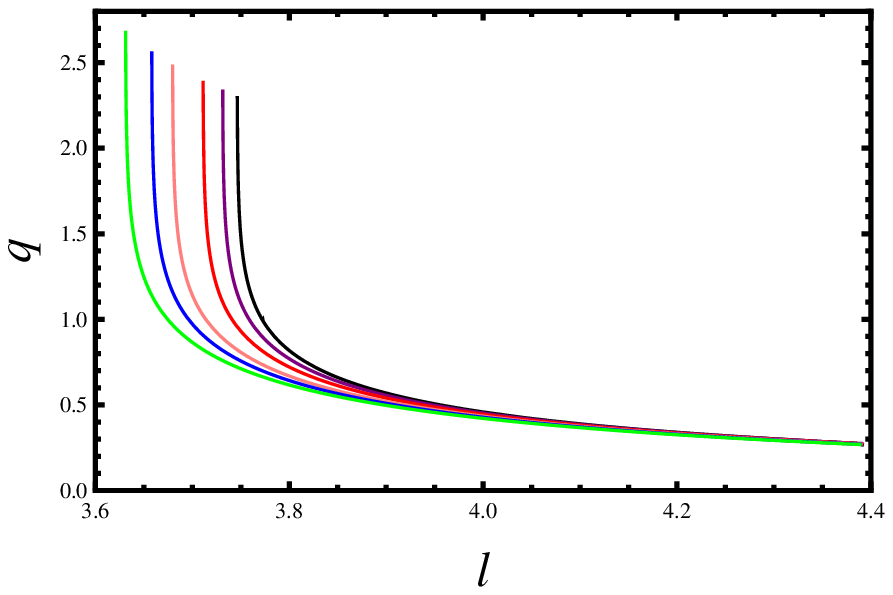}}
\subfigure[$E=0.98$]{\label{ql98}
\includegraphics[width=6cm]{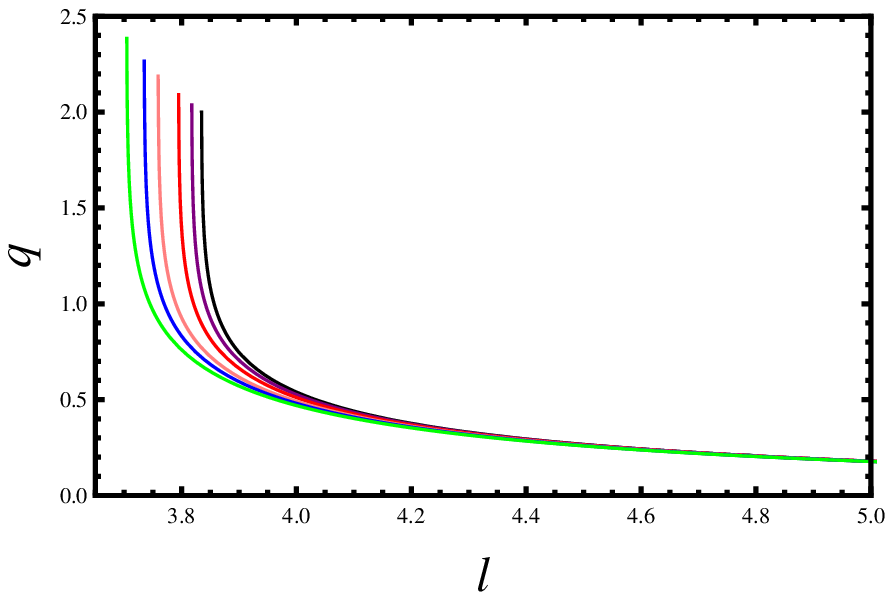}}}
\caption{$q$ vs $l$. The parameter $\omega$=0, 0.4, 0.6, 0.8, 0.9, and 1 from right to left. (a) $E=0.95$. (b) $E=0.96$. (c) $E=0.97$. (d) $E=0.98$.}\label{pql98}
\end{figure}

Following Ref. \cite{Levin}, we introduce a new parameter $q$, which is defined by
\begin{eqnarray}
 q=\frac{\Delta\phi}{2\pi}-1.
\end{eqnarray}
For a bound orbit, one can calculate $q$ with given $\omega$, $E$, and $l$. In order to show the behavior of $q$, we plot  $q$ vs. $E$ and $q$ vs. $l$ in Figs. \ref{pqe09} and  \ref{pql98}, respectively. For a bound orbit, the value of the angular momentum $l$ can vary from $l_{\text{isc}}$ to $l_{\text{mb}}$. In order to parameterize it, we express the angular momentum in the following form
\begin{eqnarray}
 l=l_{\text{isc}}+\epsilon(l_{\text{mb}}-l_{\text{isc}}).
\end{eqnarray}
Then the parameter $\epsilon$ will be limited in the range of (0, 1) for a bound orbit. For example, $\epsilon$=0 and 1 always, respectively, correspond to $l=l_{\text{isc}}$ and $l=l_{\text{mb}}$ for different values of $\omega$. Thus, no bound orbit exists when the parameter $\epsilon$ is above one. Conservation of angular momentum also implies that $\epsilon$ keeps constant. The parameter $q$ is displayed as a function of energy $E$ in Fig. \ref{pqe09} with $\epsilon$=0.3, 0.5, 0.7, and 0.9. We find from Fig. \ref{pqe09} that the parameter $q$ slowly increases with the energy $E$ at first, then when the maximal energy is approached, $q$ suddenly blows up. Note that the maximal energy decreases with $\omega$. With a detailed compare, one can also find that the maximal energy increases with $\epsilon$.

On the other hand, we show the value of $q$ in Fig. \ref{pql98} for fixed $E$=0.95, 0.96, 0.97, and 0.98. All the results reveal that $q$ decreases with the angular momentum $l$. Interestingly, $q$ goes to positive infinity at the minimum of $l$. Moreover, the minimum value increases with $E$, while decreases with $\omega$.

For a periodic orbit, the value of $q$ is a rational number, and it can be decomposed in terms of three integers $(z, w, v)$:
\begin{eqnarray}
 q=w+\frac{v}{z}.
\end{eqnarray}
In previous work \cite{Levin,Grossman,Levin2,Misra,Babar}, it was found that periodic orbits around the Schwarzschild black holes, Kerr black holes, and naked singularities are characterized by these three integers \cite{Levin,Grossman,Levin2,Misra,Babar}. As pointed out in Ref. \cite{Levin}, each integer has its own geometric interpretation. For example, integers $z$, $w$, and $v$ are, respectively, the zoom number, the whirls number, and the number of vertices formed by joining the successive apastra of the periodic orbit.

\begin{figure}
\center{
\subfigure[$E$=0.964346]{\label{Orbitsa}
\includegraphics[width=4.5cm]{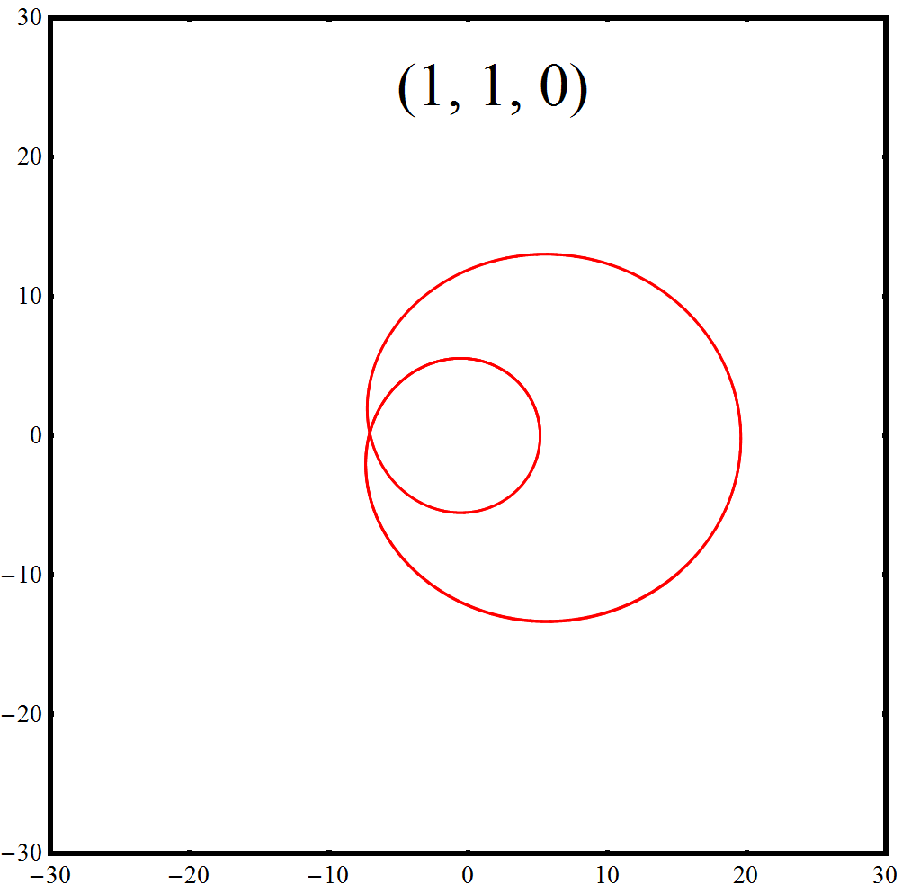}}
\subfigure[$E$=0.967744]{\label{Orbitsb}
\includegraphics[width=4.5cm]{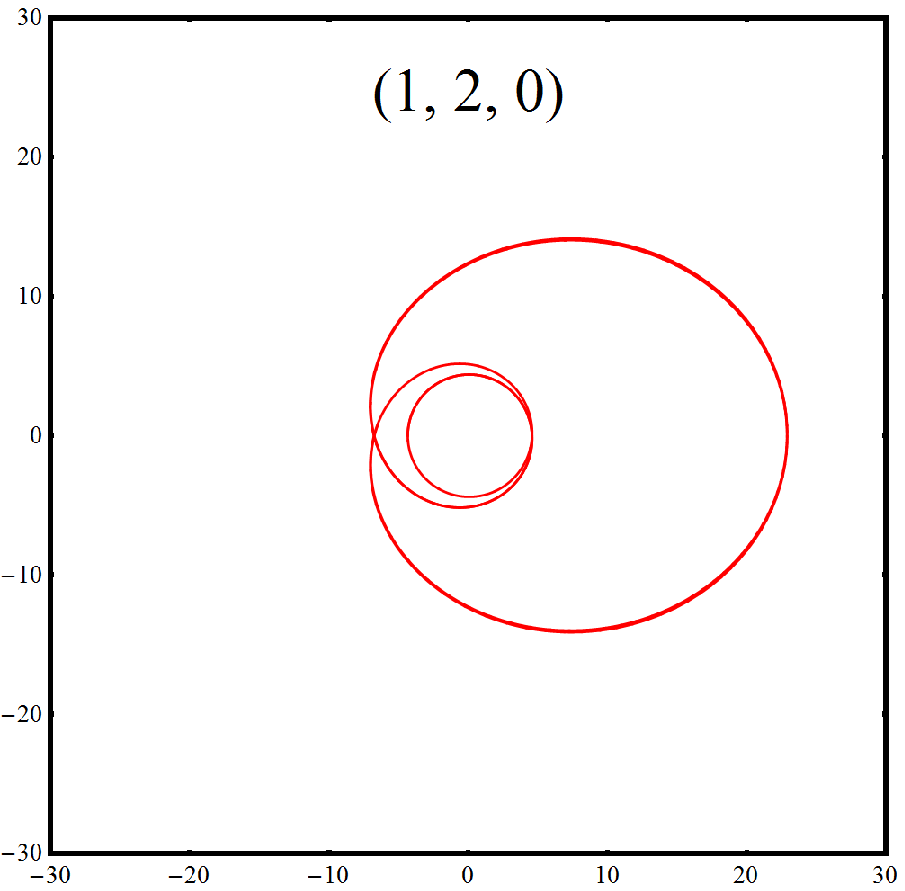}}
\subfigure[$E$=0.967822]{\label{Orbitsc}
\includegraphics[width=4.5cm]{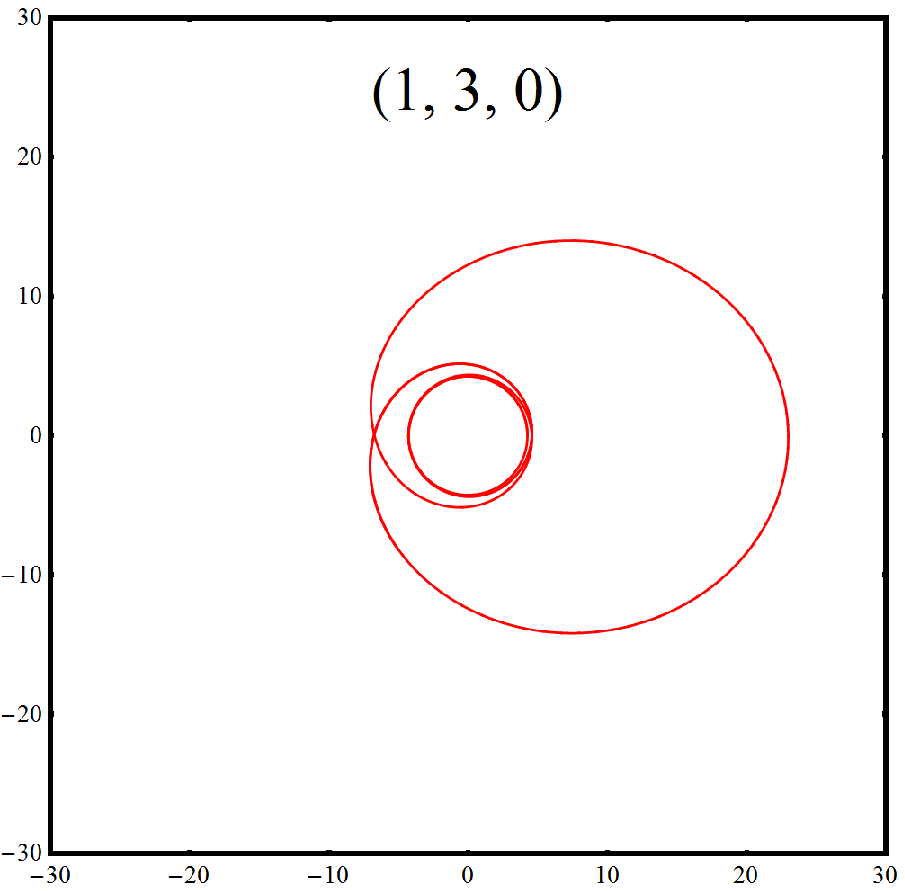}}}\\
\center{\subfigure[$E$=0.967314]{\label{Orbitsd}
\includegraphics[width=4.5cm]{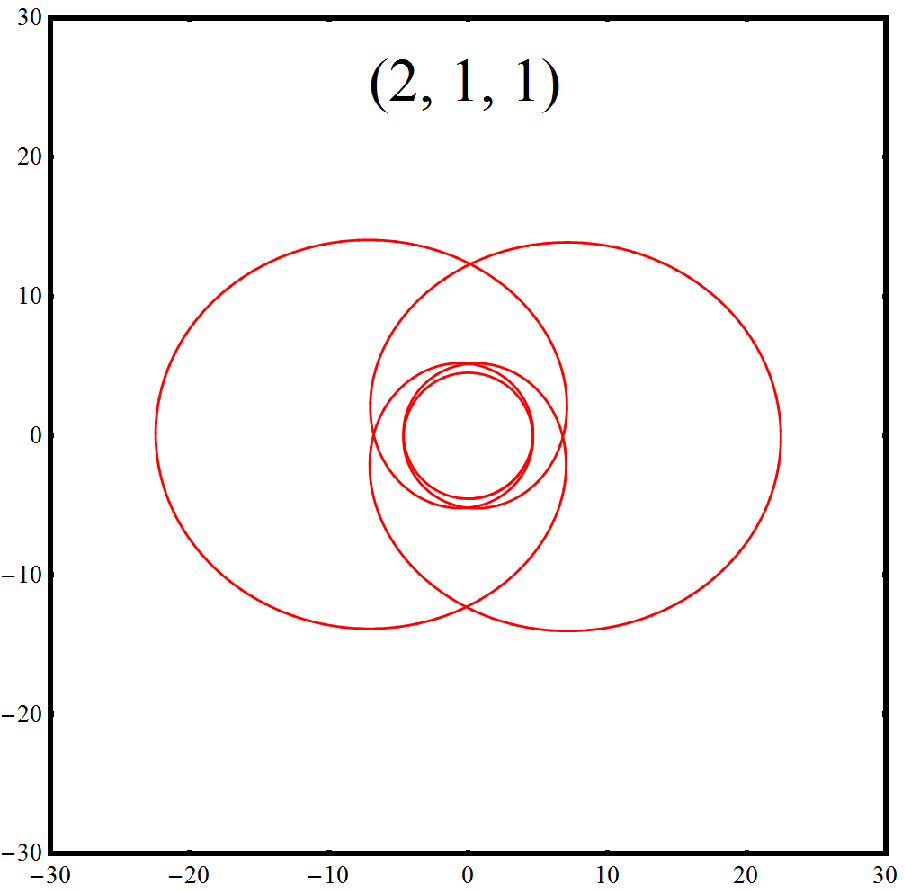}}
\subfigure[$E$=0.967811]{\label{Orbitse}
\includegraphics[width=4.5cm]{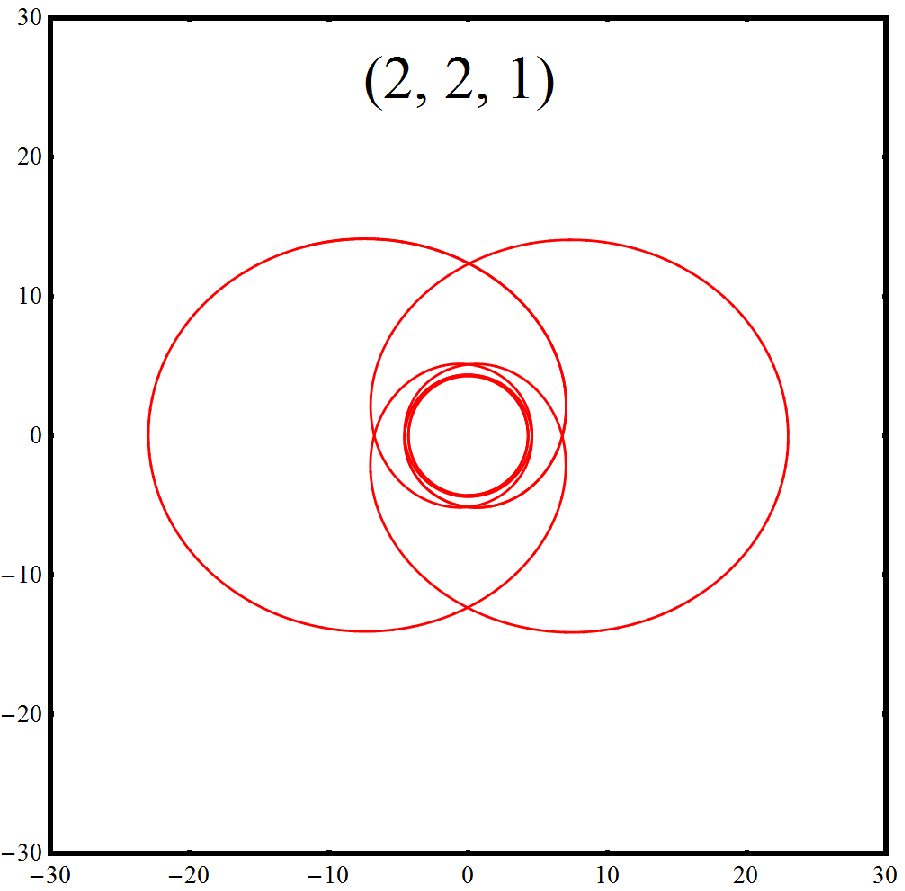}}
\subfigure[$E$=0.967823]{\label{Orbitsf}
\includegraphics[width=4.5cm]{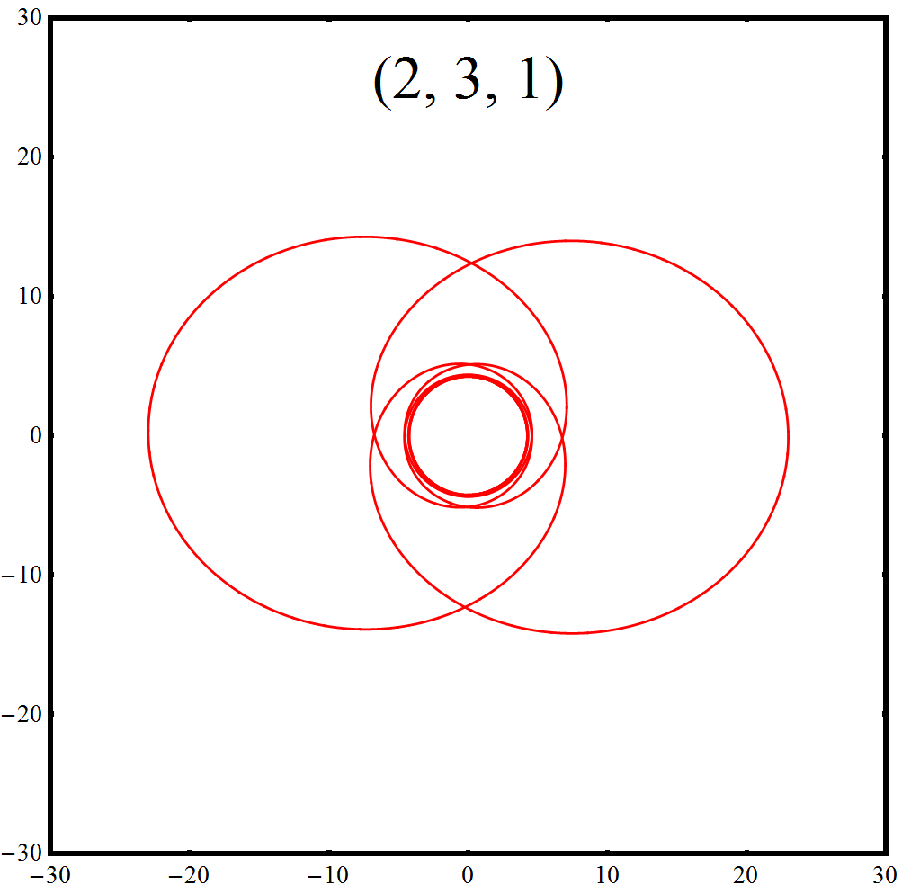}}}\\
\center{\subfigure[$E$=0.954275]{\label{Orbitsg}
\includegraphics[width=4.5cm]{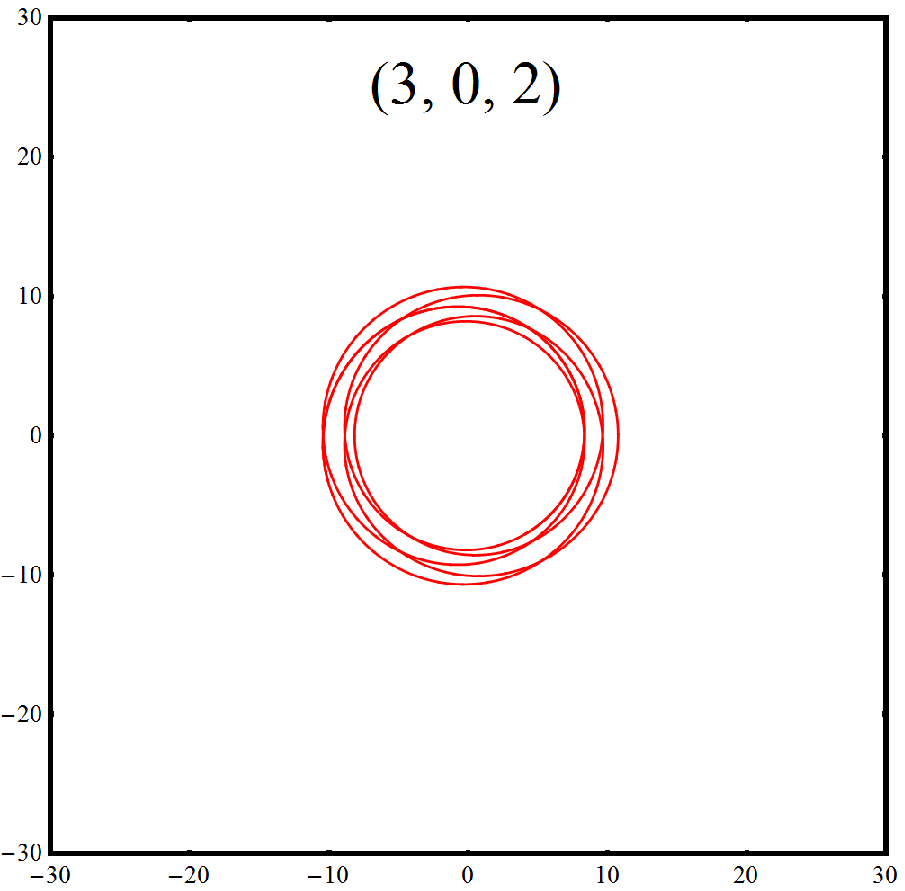}}
\subfigure[$E$=0.967551]{\label{Orbitsh}
\includegraphics[width=4.5cm]{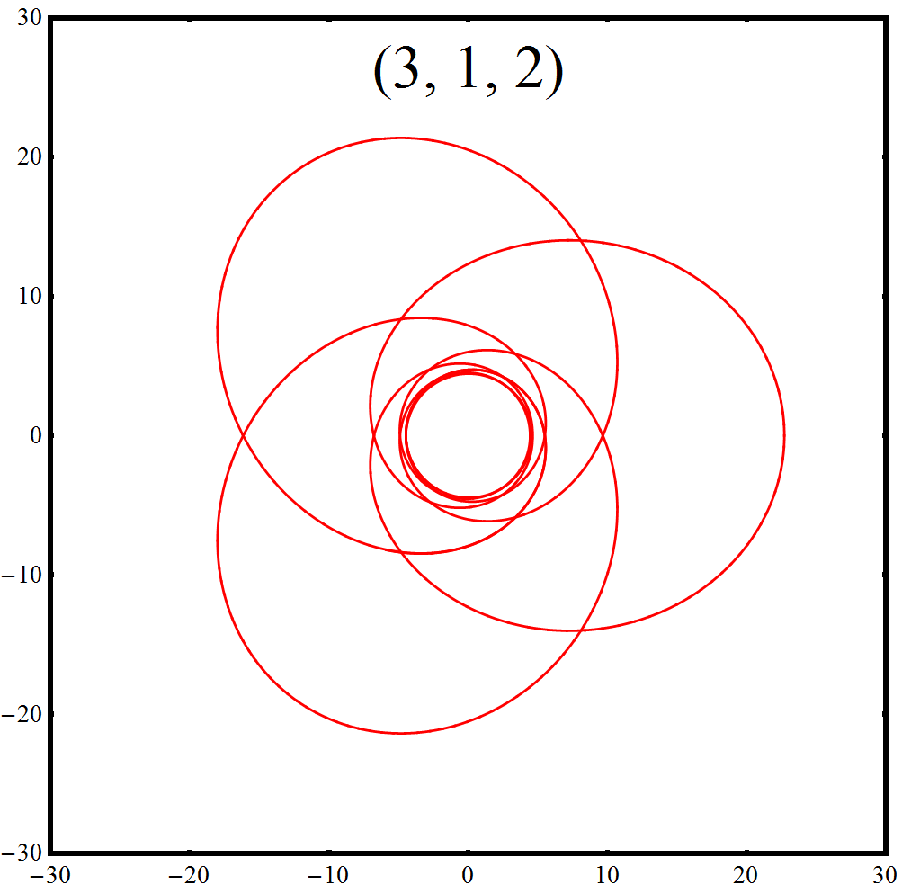}}
\subfigure[$E$=0.967817]{\label{Orbitsi}
\includegraphics[width=4.5cm]{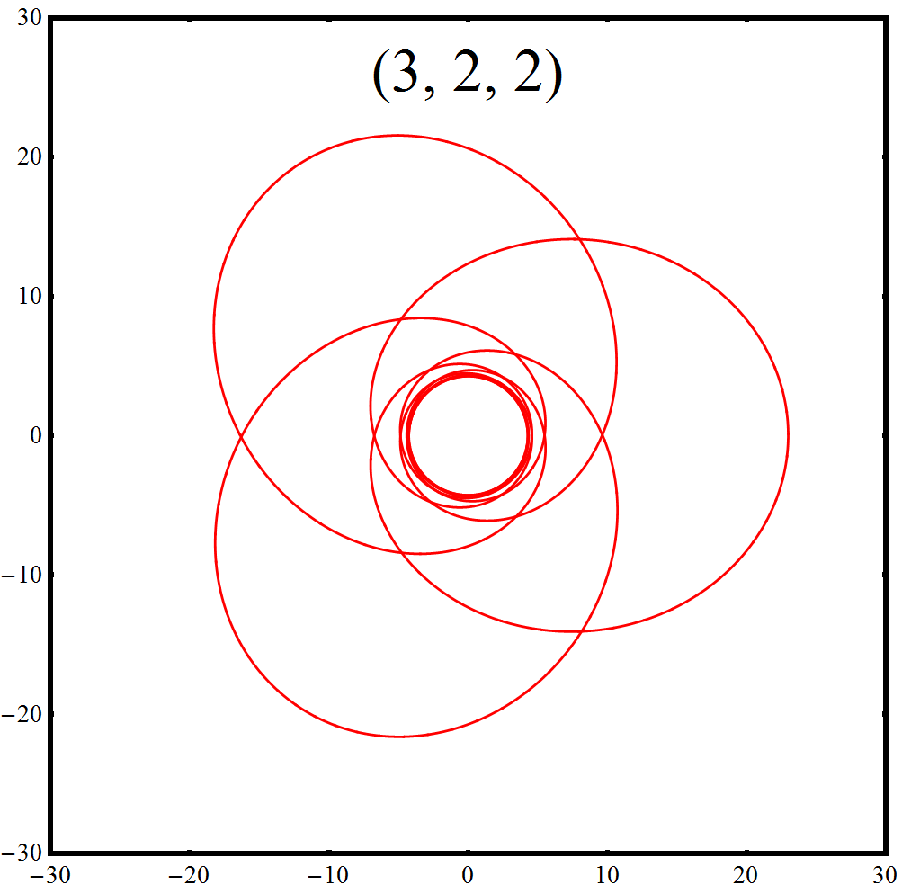}}}\\
\center{\subfigure[$E$=0.958277]{\label{Orbitsj}
\includegraphics[width=4.5cm]{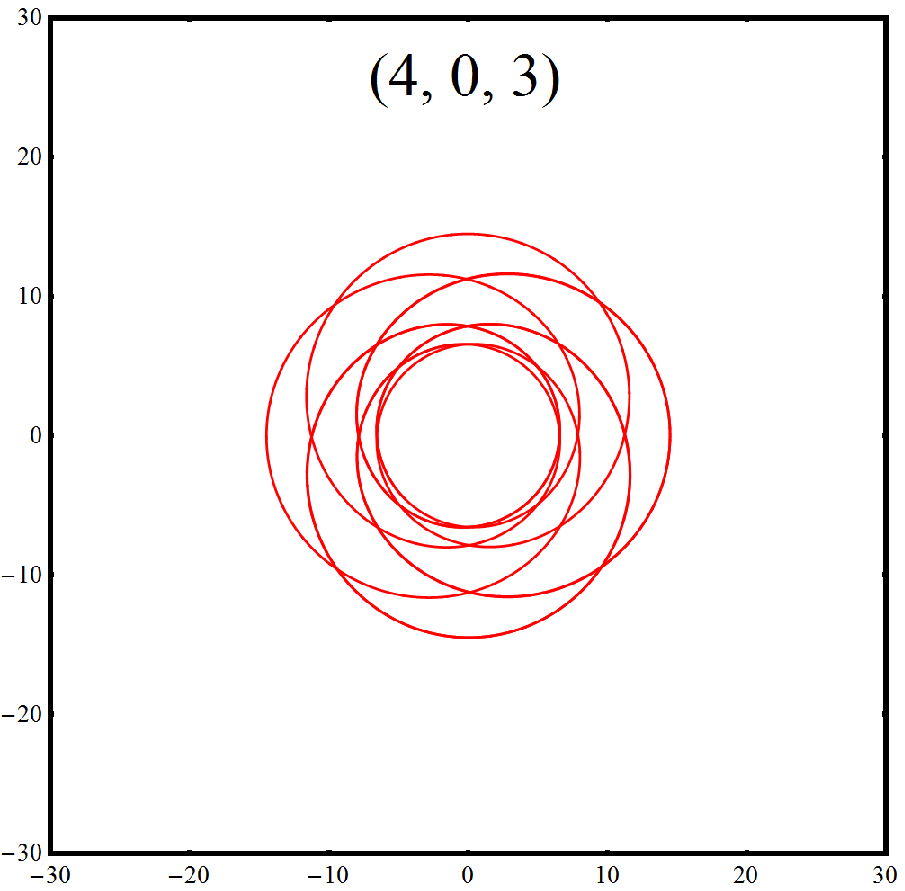}}
\subfigure[$E$=0.967624]{\label{Orbitsk}
\includegraphics[width=4.5cm]{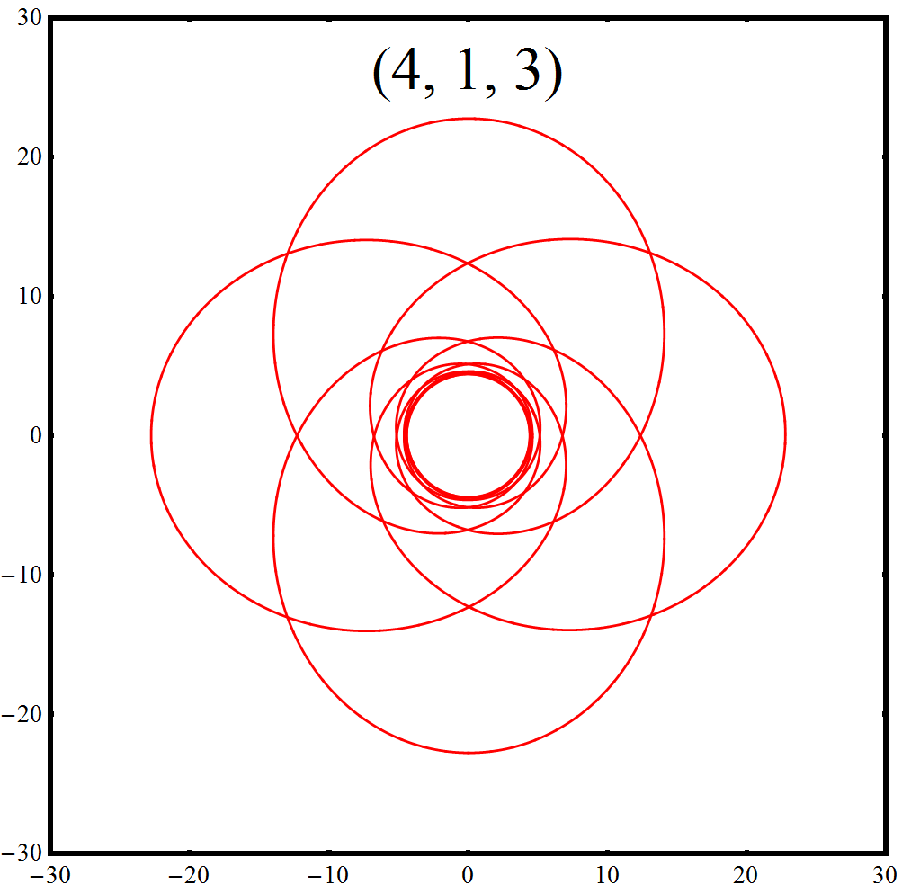}}
\subfigure[$E$=0.967819]{\label{Orbitsl}
\includegraphics[width=4.5cm]{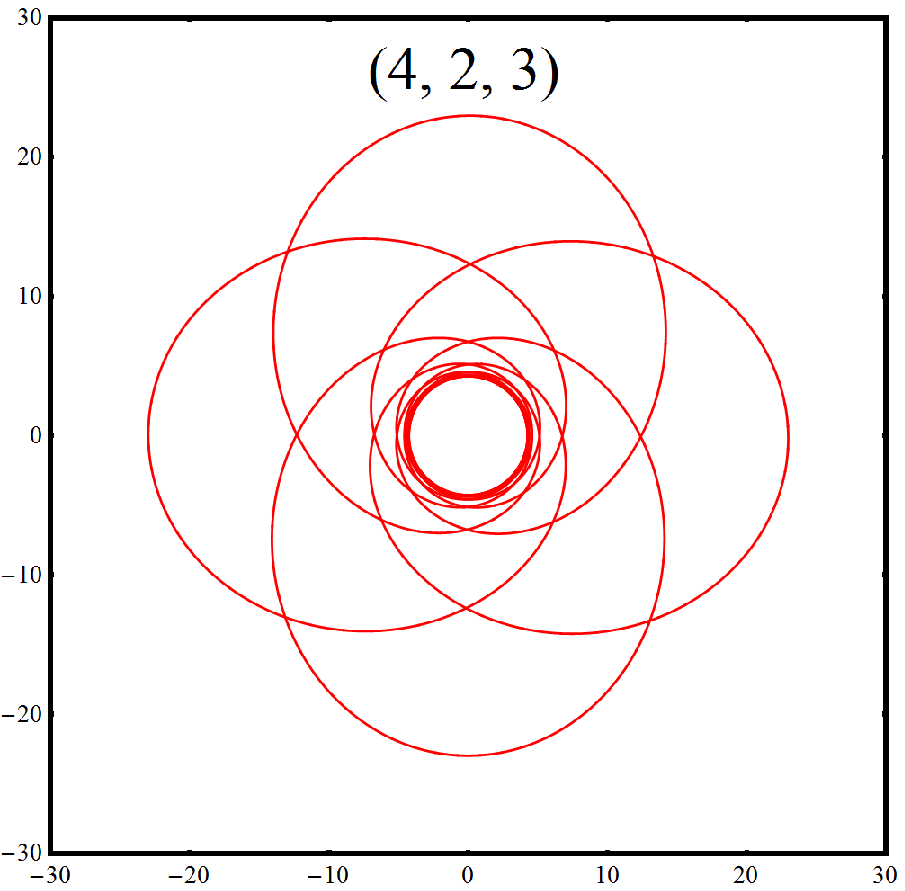}}}
\caption{Periodic orbits of different $(z, w, v)$ around a KS black hole with $\omega=0.5$ and $\epsilon=0.5$. (a) $E$=0.964346. (b) $E$=0.967744. (c) $E$=0.967822. (d) $E$=0.967314. (e) $E$=0.967811. (f) $E$=0.967823. (g) $E$=0.954275. (h) $E$=0.967551. (i) $E$=0.967817. (j) $E$=0.958277. (k) $E$=0.967624. (l) $E$=0.967819.}\label{pOrbitsa}
\end{figure}

In Fig. \ref{pOrbitsa}, we show the visualization of periodic orbits for fixed $\omega=0.5$ and $\epsilon=0.5$ for different $(z, w, v)$. Obviously, $z$ describes the number of the leaf pattern for the orbits. With the increase of $z$, the leaf pattern grows, and the orbit becomes more complicated.

\begin{table}[h]
\begin{center}
\begin{tabular}{ccccccccc}
  \hline\hline
  $\omega$ & $E_{(1,1,0)}$ & $E_{(1,2,0)}$ & $E_{(2,1,1)}$ & $E_{(2,2,1)}$
  &$E_{(3,1,2)}$ & $E_{(3,2,2)}$& $E_{(4,1,3)}$ & $E_{(4,2,3)}$\\\hline
  0  & 0.953628 & 0.957086 & 0.956607 & 0.957170
     & 0.956864 & 0.957178 & 0.956946 & 0.957181\\\hline
  0.2& 0.953430 & 0.956953 & 0.956462 & 0.957040
     & 0.956725 & 0.957048 & 0.956809 & 0.957051\\\hline
  0.4& 0.952802 & 0.956537 & 0.956007 & 0.956633
     & 0.956290 & 0.956643 & 0.956380 & 0.956646\\\hline
  0.6& 0.951635 & 0.955784 & 0.955173 & 0.955899
     & 0.955496 & 0.955910 & 0.955600 & 0.955914\\\hline
  0.8& 0.949666 & 0.954565 & 0.953800 & 0.954716
     & 0.954198 & 0.954736 & 0.954329 & 0.954741\\\hline
  1.0& 0.946222 & 0.952564 & 0.951467 & 0.952815
     & 0.952023 & 0.952836 & 0.952212 & 0.952854\\\hline\hline
\end{tabular}
\caption{The energy $E$ for the orbits with different $(z, w, v)$ and different black hole parameter $\omega$. The angular momentum parameter $\epsilon=0.3$. Note that $\omega$=0 denotes the Schwarzschild black hole case.}\label{tab1}
\end{center}
\end{table}
\begin{table}[h]
\begin{center}
\begin{tabular}{ccccccccc}
  \hline\hline
  $\omega$ & $E_{(1,1,0)}$ & $E_{(1,2,0)}$ & $E_{(2,1,1)}$ & $E_{(2,2,1)}$
  &$E_{(3,1,2)}$ & $E_{(3,2,2)}$& $E_{(4,1,3)}$ & $E_{(4,2,3)}$\\\hline
  0  & 0.965425 & 0.968383 & 0.968026 & 0.968434
     & 0.968225 & 0.968438 & 0.968285 & 0.96844\\\hline
  0.2& 0.965265 & 0.968286 & 0.967919 & 0.96834
     & 0.968123 & 0.968344 & 0.968185 & 0.968345\\\hline
  0.4& 0.964757 & 0.967984 & 0.967583 & 0.968045
     & 0.967804 & 0.968050 & 0.967872 & 0.968051\\\hline
  0.6& 0.963802 & 0.967434 & 0.966963 & 0.967511
     & 0.967222 & 0.967518 & 0.967302 & 0.96752\\\hline
  0.8& 0.962156 & 0.966538 & 0.965930 & 0.966654
     & 0.966261 & 0.966664 & 0.966366 & 0.966667\\\hline
  1.0& 0.959159 & 0.965072 & 0.964126 & 0.965264
     & 0.964617 & 0.965284 & 0.964780 & 0.965291\\\hline\hline
\end{tabular}
\caption{The energy $E$ for the orbits with different $(z, w, v)$ and different black hole parameter $\omega$. The angular momentum parameter $\epsilon=0.5$.}\label{tab2}
\end{center}
\end{table}

Taking $\epsilon=$0.3 and 0.5, we also list the corresponding energy for each periodic orbit in Tables \ref{tab1} and \ref{tab2}. Since the Schwarzschild black hole corresponds to $\omega$=0, we can find that, for each periodic orbit with fixed $(z, w, v)$, the particle orbiting a Schwarzschild black hole always has the highest energy. The energy decreases with the parameter $\omega$. When the extremal black hole $\omega$=1 is approached, the energy reaches its minimum. Moreover, the energy of the particle grows with the parameter $\epsilon$. For example, for a black hole with $\omega$=0.6, $E_{(2,1,1)}$=0.955173 and 0.966963 for $\epsilon$=0.3 and 0.5, respectively.

In general, accompanied by the emission of gravitational waves, the energy and angular momentum of the orbiting particle decrease. So there will be a transition for the orbit from one energy level diagram with given $l$ to another one with a lower $l$. As shown above, the parameter $q$ closely depends on both the energy $E$ and the angular momentum $l$, so the rate of change of q is
\begin{eqnarray}
 \frac{dq}{dt}=\frac{\partial q}{\partial E}\frac{dE}{dt}
               +\frac{\partial q}{\partial l}\frac{dl}{dt}.
\end{eqnarray}
During the successive orbit transition in the inspiral stage, there may exist the resonances satisfying $dq/dt\approx0$. From Figs. \ref{pqe09} and \ref{pql98}, one can find that $\partial q/\partial E>$0 and $\partial q/\partial l<$0. Thus, it is possible to satisfy condition $dq/dt\approx0$.

\begin{figure}
\center{
\includegraphics[width=8cm]{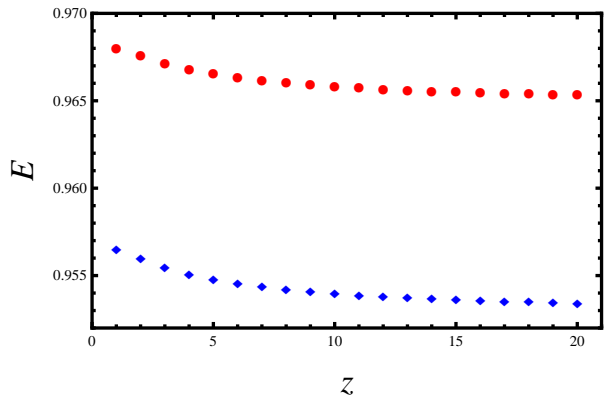}}
\caption{The energy $E$ as a function of $z$ for the orbits $(z, 1, 1)$ with the black hole parameter $\omega$=0.5. The angular momentum is fixed with $\epsilon$=0.3 (bottom) and 0.5 (top).}\label{Eezz}
\end{figure}

On the other hand, the energy of the orbiting particle gradually decreases with emitting gravitational waves. Supposing the angular momentum $l$ keeps constant, we plot the energy $E$ as a function of $z$ for the periodic orbits of $(z, 1, 1)$ in Fig. \ref{Eezz}. As mentioned above, the value $z$ denotes the number of the leaf for the orbits. Interestingly, with fixed $l$, the energy decreases with $z$, which implies that the orbit will approach to an orbit with high $z$ by decreasing particle energy. As expected, the angular momentum $l$ will also be radiated away in the form of gravitational waves. From Fig. \ref{Eezz}, we can also find that these particles have lower energy for small $l$ or $\epsilon$ by comparing the cases of $\epsilon$=0.5 and 0.3. Thus, the orbit of a particle with emitting gravitational waves will finally tend to the one with lower energy and angular momentum while with high $z$. It is also worthwhile noting that with $z\rightarrow\infty$, $q$ will approach an integer, and thus a circular orbit, specifically the ISCO, is formed.

\section{Conclusions}
\label{Conclusion}

In the present work, we studied the property of the periodic orbits around a KS black hole in the deformed HL gravity. At first, we obtained the geodesics for the particle moving in the KS black hole backgrounds, which is quite different from that of the Schwarzschild black hole, while the result for the Schwarzschild case can be recovered with $\omega$=0.

Then we reexpressed the $r$-motion with an effective potential. Employing the effective potential, we numerically calculated the marginally bound orbits and ISCOs. The results show that for the marginally bound orbits, both the radius and angular momentum decrease with $\omega$. The energy and angular momentum for the ISCOs also decrease with $\omega$.

Based on the properties of the marginally bound orbits and ISCOs, we considered the periodic orbits in the KS black holes. The quantity $q$ describing the apsidal angle increases with the particle energy, while decreases with the angular momentum. In particular, for fixed $E$, $q$ increases with $\omega$. However, for fixed $l$, $q$ decreases with $\omega$.

According to Ref. \cite{Levin}, each periodic orbit is characterized by a set of parameters $(z, w, v)$. For the same orbit of constant $(z, w, v)$, the energy gets lower and lower with the increase of $\omega$. Thus, these periodic orbits in a KS black hole always have a lower energy than that of a Schwarzschild black hole. Moreover, the minimum energy is approached for the extremal KS black hole. We also expanded the study to these orbits with the same $w$ and $v$. It is exhibited that the energy decreases with $z$, which means that, with fixed angular momentum, these orbits with high $z$ generally have lower energy. When $z\rightarrow\infty$, the orbits tend to circular ones. So these circular orbits, especially the ISCOs, have lowest energy. These results may provide us a possible observational signature by testing these periodic orbits around a central source to distinguish a KS black hole from a Schwarzschild one. Furthermore, since the orbits pass a series of periodic orbits during the inspiral stage, which can be taken as transient states, these periodic orbits with non-vanishing $dq/dt$ are very important for the detection of the gravitational waves.

\section*{Acknowledgements}
This work was supported by the National Natural Science Foundation of China (Grants No. 11675064, No. 11875151, and No. 11522541) and the Fundamental Research Funds for the Central Universities (Grants No. lzujbky-2018-k11).

\end{document}